\documentclass[prb,twocolumn,amsmath,amssymb]{revtex4}

\usepackage{graphicx}% Include figure files
\usepackage{bm}% bold math
\usepackage{gensymb}
\usepackage{multirow}
\usepackage{xcolor}
\usepackage{lipsum}
\usepackage{color}
\usepackage{hyperref} % For hyperlinks in the PDF
\usepackage{newfloat}
\usepackage{siunitx}

\begin{document}

\title{Low-carrier density and fragile magnetism in a Kondo lattice system}

\author{Binod~K.~Rai$^1$}
\author{Iain~W.~H.~Oswald$^2$}
\author{Wenjing~Ban$^3$}
\author{C.-L.~Huang$^1$}
\email[]{clh@rice.edu}
\author{V.~Loganathan$^1$}
\author{A.~M.~Hallas$^{1,4}$}
\author{M.~N.~Wilson$^4$}
\author{G.~M.~Luke$^{4,5}$}
\author{L.~Harriger$^6$}
\author{Q.~Huang$^6$}
\author{Y.~Li$^1$}
\author{Sami~Dzsaber$^7$}
\author{Julia~Y.~Chan$^2$}
\author{N.~L.~Wang$^3$}
\author{Silke~Paschen$^7$}
\author{J.~W.~Lynn$^6$}
\author{Andriy~H.~Nevidomskyy$^1$}
\author{P.~Dai$^1$}
\author{Q.~Si$^1$}
\author{E.~Morosan$^1$}
\email[]{emorosan@rice.edu}

\affiliation{$^1$Department of Physics and Astronomy and Rice Center for Quantum Materials, Rice University, Houston, TX 77005, United States\\$^2$ Department of Chemistry, University of Texas at Dallas, Richardson, Texas 75080, United States\\$^3$ International Center for Quantum Materials, School of Physics, Peking University, China\\$^4$ Department of Physics and Astronomy, McMaster University, Hamilton, Ontario L8S 4M1, Canada\\$^5$Canadian Institute for Advanced Research, 180 Dundas Street West, Toronto, Ontario M5G 1M1, Canada\\$^6$NIST Center for Neutron Research, National Institute of Standards and Technology, Gaithersburg, MD 20899, United States\\$^7$ Institute of Solid State Physics, Vienna University of Technology, Wiedner Hauptstr. 8-10, 1040 Vienna, Austria}

\date{\today}
\begin{abstract}
Kondo-based semimetals and semiconductors are of extensive current interest as a viable platform for strongly correlated states in the dilute carrier limit. It is thus important to understand the routes to understand such systems. One established pathway is through Kondo effect in metallic non-magnetic analogues, in the so called half-filling case of one conduction electron and one 4$f$ electron per site. Here we advance a new mechanism, through which Kondo-based semimetals develop out of conduction electrons with a low-carrier density in the presence of an even number of rare-earth sites. We demonstrate this effect by studying the Kondo material Yb$_3$Ir$_4$Ge$_{13}$ along with its closed-4$f$-shell counterpart, Lu$_3$Ir$_4$Ge$_{13}$. Through magnetotransport, optical conductivity and thermodynamic measurements, we establish that the correlated semimetallic state of Yb$_3$Ir$_4$Ge$_{13}$ below its Kondo temperature originates from the Kondo effect of a low-carrier conduction-electron background. In addition, it displays fragile magnetism at very low temperatures, which, in turn, can be tuned to a non Fermi liquid regime through Lu-for-Yb substitution. These findings are connected with recent theoretical studies in simplified models. Our results open an entirely new venue to explore the strong correlation physics in a semimetallic environment.
\end{abstract}
\maketitle

The proximity of the \textit{f} energy level to the Fermi energy $E_{\rm F}$ in rare earth compounds often causes the hybridization of the local moments with the conduction electrons. The resulting Kondo effect gives rise to a broad range of electronic properties, from metallic heavy fermion (HF) systems to Kondo insulators, and intermediate low-carrier Kondo semiconductors or semimetals. Kondo systems have garnered much attention in recent years because of their exotic electronic and magnetic behavior, including quantum criticality, breakdown of the Fermi-liquid picture, unconventional superconductivity, and topologically-protected states \cite{Stewart,Pfleiderer2009,Takabatake1998,SiSteglich}. Only a small number of 4$f$ low-carrier HF systems have been reported so far, e.g., CeNiSn and CeRhSb \cite{Takatabake1996,Nishigori1996}, CeNi$_{2-\delta}$As$_2$ \cite{Luo and Thompson}, and Ce$_3$Bi$_4$Pd$_3$ \cite{Dzsaber}. These compounds realize HF physics in an entirely new parameter regime which is away from the canonical ``half filling'' case, and thus new low-carrier Kondo systems are called for to broaden our knowledge in such a regime.

Our recent discovery of the six non-magnetic germanides Y$_3$T$_4$Ge$_{13-x}$ (T = Ir, Rh, Os) and Lu$_3$T$_4$Ge$_{13-x}$ (T = Co, Rh, Os) \cite{Rai2015} pointed to a possible generalization of low-carrier behavior in these ``3-4-13" germanides, due to the negative temperature coefficients of the electrical resistivity $\rho(T)$, \textit{i.e.}, d$\rho$/d$T <$ 0, compared to the normal metal behavior (d$\rho$/d$T >$ 0) in the stannide analogues \cite{Oswald2017}. Here we report the discovery that the correlated semimetal Yb$_3$Ir$_4$Ge$_{13}$ (YbIG) \cite{Binod2016} emerges from Kondo effect in a nonmagnetic dilute-carrier system, Lu$_3$Ir$_4$Ge$_{13}$ (LIG). YbIG exhibits d$\rho$/d$T <$ 0 in the whole measured temperature range from $300$ to $0.1$ K, including inside a fragile magnetic state below $T^{\ast}_{mag}=$ 0.9 K. The underlying electronic properties of YbIG are dictated by its non-magnetic analogue LIG, which, remarkably, also shows semimetal\textit{-like} behavior between 300 and 2.8 K, below which LIG becomes superconducting. This is in stark contrast with the metallic behavior found in the non-magnetic analogues of other known Kondo semimetals \cite{Malik1995,Echizen1999,Luo2012,Tafti2015,Sun2016}. Optical conductivity measurements and band structure calculations reinforce the semimetal-like nature of YbIG and LIG. Moreover, the substitution of non-magnetic Lu on the Yb site induces non-Fermi-liquid (nFL) behavior, as evidenced by the divergence of the magnetic susceptibility $M/H$ and specific heat $C_p/T$ at low $T$. The coexistence of the low-carrier density, Kondo effect and the associated semimetal behavior, and fragile magnetism makes YbIG one of the most complex $f$ electron systems. We rationalize our finding using a recently developed framework of the Kondo-lattice effect in the dilute-carrier limit \cite{Feng and Si}.

\begin{figure}[htbp]
	\includegraphics[clip,width=\columnwidth]{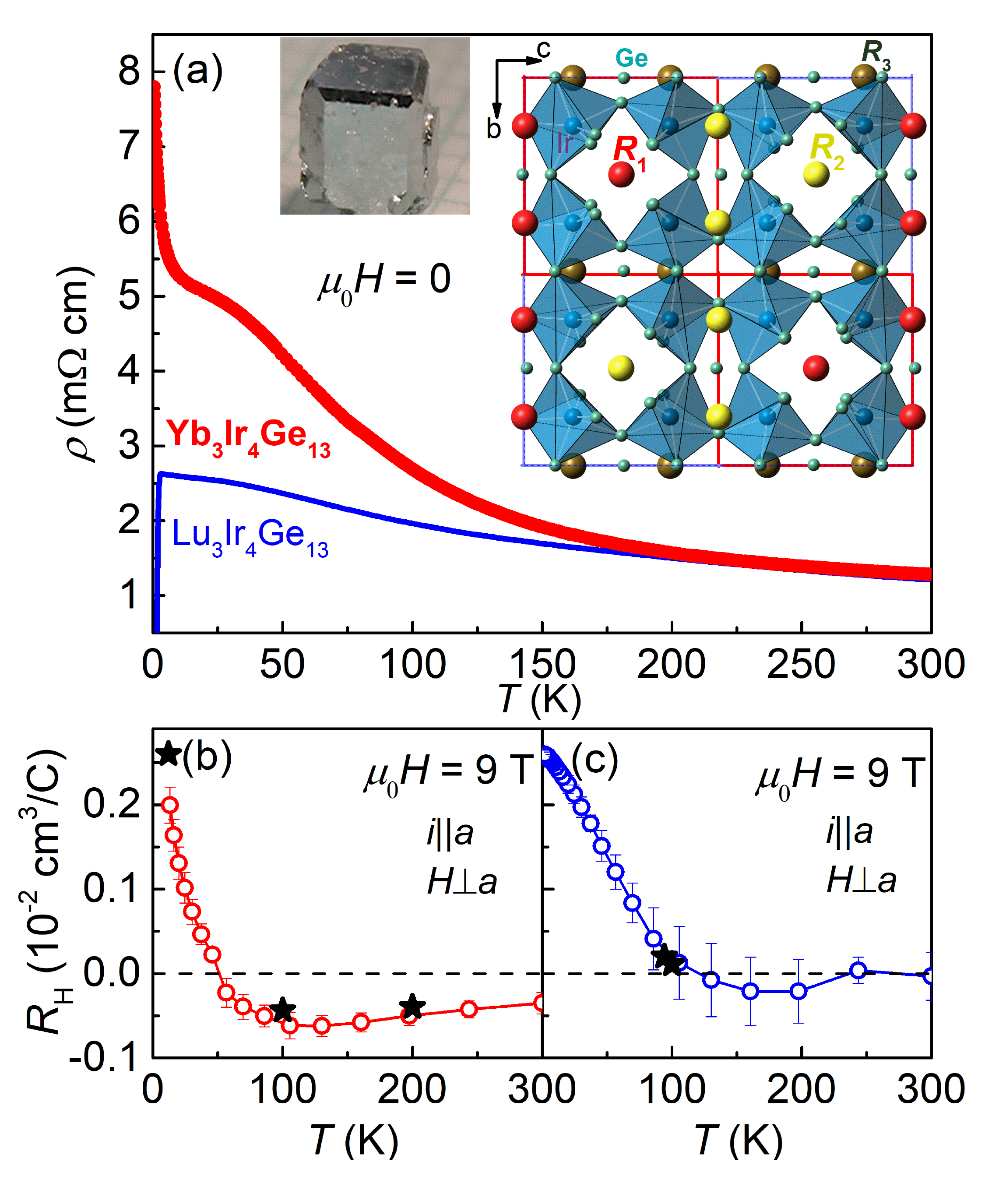}
\caption{\label{Fig1} (a) Zero field temperature-dependent electrical resistivity $\rho$ of $R{_3}$Ir${_4}$Ge$_{13}$, $R$ = Yb and Lu. A sudden drop of $\rho(T)$ in Lu${_3}$Ir${_4}$Ge$_{13}$ is due to a superconducting transition at 2.8 K. The inset shows the crystal structure of $R{_3}$Ir${_4}$Ge$_{13}$ and a picture of a crystal. (b) and (c) show temperature-dependent Hall coefficients $R_{\rm H}(T)$ measured at $\mu_0 H$ = 9 T for Yb${_3}$Ir${_4}$Ge$_{13}$ and Lu${_3}$Ir${_4}$Ge$_{13}$, respectively. Black stars are data measured in a field sweep mode at a constant $T$. The error bar represents a difference between cooling and warming measurements.}
\end{figure}

The growth of single crystals of YbIG and LIG is described in the Supplementary Materials, with the room-temperature powder diffraction pattern refinement results shown in Supplementary Figure~S1. We confirmed that both YbIG and LIG adopt the same tetragonal space group $I4_{1}/amd$. The temperature-dependent magnetization $M/H$ for YbIG \cite{Binod2016} is similar to many other 4$f$ magnetic systems with comparable Kondo temperatures $T_{\rm K} \approx 3.5$ K \cite{Yang2008}: Curie-Weiss behavior is evidenced at high temperatures by a linear inverse susceptibility $H/M$, with a peak in $M/H$ suggesting antiferromagnetic correlations below $\sim$ 1 K. However, other 4$f$ magnetic systems are usually metallic \cite{Stewart}, while the electrical resistivity $\rho$ of YbIG (red symbols, Fig.~\ref{Fig1}(a)) shows non-metallic behavior even at high temperatures ($T \gg T_{\rm K}$). This cannot be due to incoherent scattering of the conduction electrons by localized $f$ moments, since non-magnetic LIG exhibits similar d$\rho$/d$T <$ 0 behavior (blue line, Fig. \ref{Fig1}(a)). Upon cooling below 50 K, $\rho(T)$ increases further, and a signature of Kondo effect appears in YbIG as a shoulder near 40 K. In LIG a monotonic increase of $\rho(T)$ is observed down to 2.8 K, below which LIG becomes superconducting (Supplementary Figure~S2). Similar semiconducting $\rho(T)$ behavior in LIG has recently also been reported by Kumar $et$ $al.$ \cite{Kumar2018}. However, they grew the crystal by using the Czochralski method and reported a different crystal structure $Pm\bar{3}n$. This reflects that reaction conditions and synthesis methods could lead to slightly different polymorphs.

\begin{figure}[htbh]
  \includegraphics[clip,width=\columnwidth]{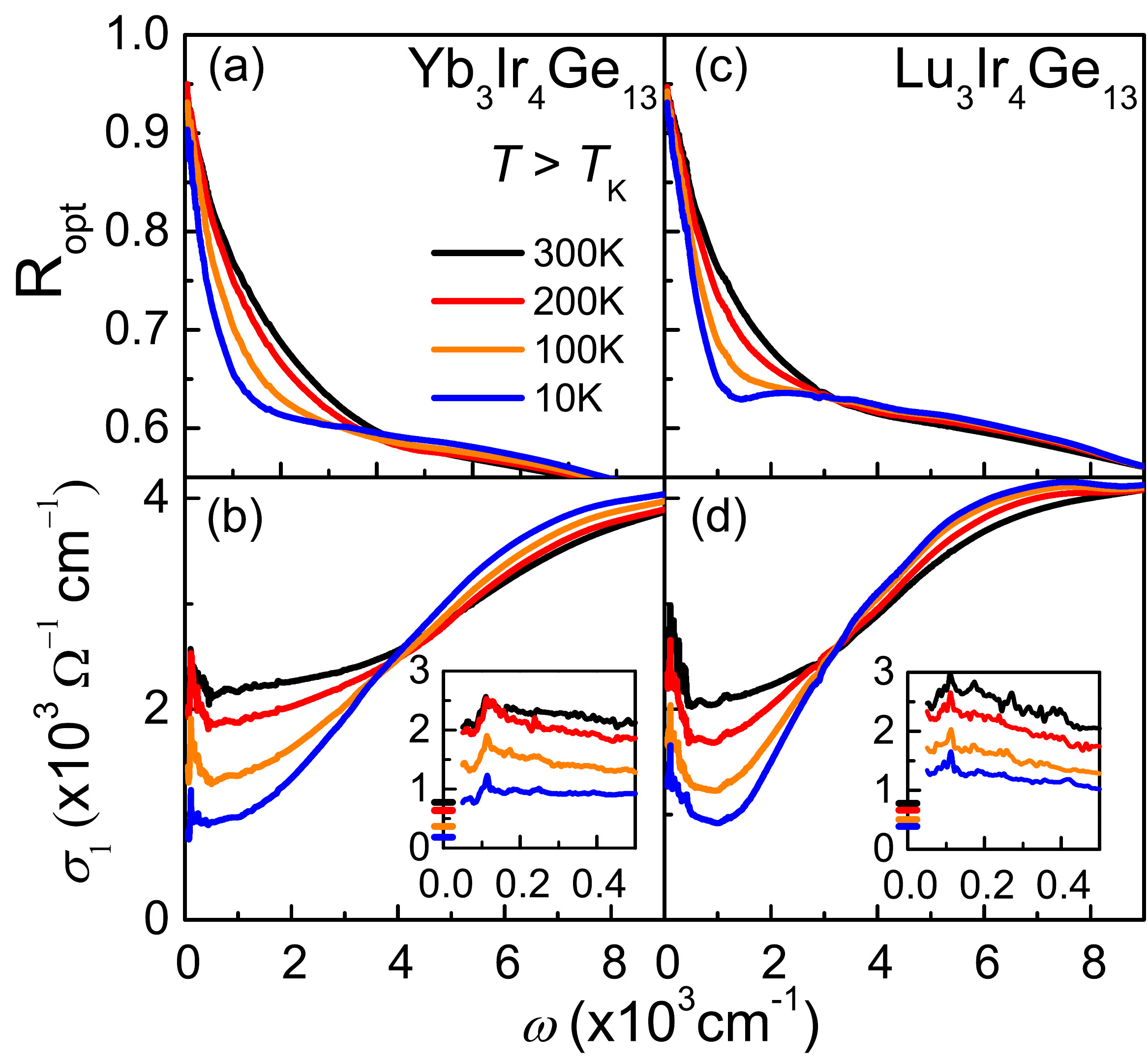}
\caption{\label{Fig2} Optical reflectance $R_{\rm {opt}}$ and optical conductivity $\sigma_1$ as a function of frequency $\omega$ of Yb$_3$Ir$_4$Ge$_{13}$ (a,b) and Lu$_3$Ir$_4$Ge$_{13}$ (c,d) at different temperatures $T$ = 10, 100, 200 and 300 K. The insets in (b) and (d) highlight the low frequency part of $\sigma_1$ and ticks at zero frequency mark the independently measured dc conductivity data (from $\rho (T)$) at the corresponding temperatures.}
\end{figure}

The low-carrier nature in YbIG and LIG is verified by temperature-dependent Hall coefficient data $R_{\rm H}(T)$, shown in Fig.~\ref{Fig1} (b) and (c), respectively. The data for YbIG are shown down to 16 K (above $T_{\rm K}$), below which the anomalous Hall effect dominates, while the data for LIG are shown down to 2 K. The Hall resistance $\rho_{\rm H}(H)$ was measured at several temperatures and the $R_{\rm H}$ values, obtained from linear fits, (black stars in Fig.~\ref{Fig1} (b) and (c)) are consistent with the $R_{\rm H}(T)$ data. A sign change of $R_{\rm H}(T)$ indicates a change in carrier type from electrons at high temperatures to holes at low temperatures for both systems. Upon further cooling below the zero crossing, the carrier density $n$ in a single-band picture ($n$ = 1/e$R_{\rm H}$) decreases, consistent with the optical measurements shown in Fig. \ref{Fig2}. Surprisingly, at the lowest temperatures, the estimated $n$ values for YbIG and LIG are comparable and lower than those of Kondo metals, $n \approx 3 \times 10^{21}$ cm$^{-3}$.  Interestingly, also the isostructural compound Y${_3}$Ir${_4}$Ge$_{13}$ exhibits low-carrier behavior with $n \approx 6.67 \times 10^{20}$ cm$^{-3}$ at 50 K \cite{Strydom2007}. In other Kondo semiconductors, the formation of a narrow hybridization gap, identified for instance by optical conductivity measurements, is deemed responsible for the low-carrier density \cite{Matsunami2002}. 

Such a gap feature, however, is not seen in our optical data (Fig.~\ref{Fig2}). Overall, similar optical properties are registered for YbIG (Fig. \ref{Fig2} (a) and (b)) and LIG (Fig. \ref{Fig2} (c) and (d)), in both the optical reflectance $R_{\rm {opt}}$($\omega$) and the real part of the optical conductivity $\sigma_1(\omega)$. Note that all data for YbIG are taken at $T > T_{\mathrm K}$. Both compounds show nominally metallic frequency dependence at all measured temperatures: $R_{\rm {opt}}$($\omega$) at low energy increases rapidly with decreasing frequency and approaches unity in the zero frequency limit. As a result, a reflectance edge, though overdamped, is seen in the measured $R_{\rm {opt}}$($\omega$) below 4000 cm$^{-1}$, and a Drude-like peak is seen at low frequency in the $\sigma_1(\omega)$ spectra. For both compounds, the low-frequency reflectance $R_{\rm {opt}}$($\omega$) values decrease with decreasing temperature (Fig. \ref{Fig2} (a) and (c)), leading to a drop of the conductivity in the low-frequency regime. The temperature dependence of $\sigma_1(\omega)$ for YbIG and LIG is consistent with the semimetal-like evolution of $\rho(T)$. A striking observation is that the reflectance edge is located at a rather low energy ($\omega = 3 - 4 \times 10^{3}$ cm$^{-1}$) compared with ordinary metals, indicating that both compounds have very low plasma frequencies or carrier densities. The estimated values of the plasma frequency at 10 K are $4.8\times10^{3}$ cm$^{-1}$ for YbIG and $5.7\times10^{3}$ cm$^{-1}$ for LIG, which correspond to small carrier densities $n \approx 2.6 \times 10^{20}(m$*$/m_{e}$) cm$^{-3}$ for YbIG and $\approx 3.6 \times 10^{20}(m$*$/m_{e}$) cm$^{-3}$ for LIG, where $m$* is the effective mass and $m_{e}$ is the free electron mass. Furthermore, the low-frequency spectral weight is suppressed with decreasing temperature, leading to the formation of a minimum of $\sigma_1$ with an energy scale of about 1000 cm$^{-1}$ for both compounds. 

We note that the values of $n$ estimated from the optical conductivity data at 10~K are one order of magnitude smaller than those obtained from the Hall effect measurements at low temperatures (Fig.~\ref{Fig2}(b) and (c)). This discrepancy reflects the combination of the strong dependence of $n$ on $m$*, with the latter being expected to be enhanced in YbIG as evidenced by the large $C_p/T$ value towards zero temperature (see Fig.~\ref{Fig3}(a)), and the use of a simple single-band model for estimating $n$ from the Hall effect measurements. Still, both types of measurements point to the dilute-carrier nature of YbIG and LIG. 

\begin{figure}[th]
	\includegraphics[width=1\columnwidth]{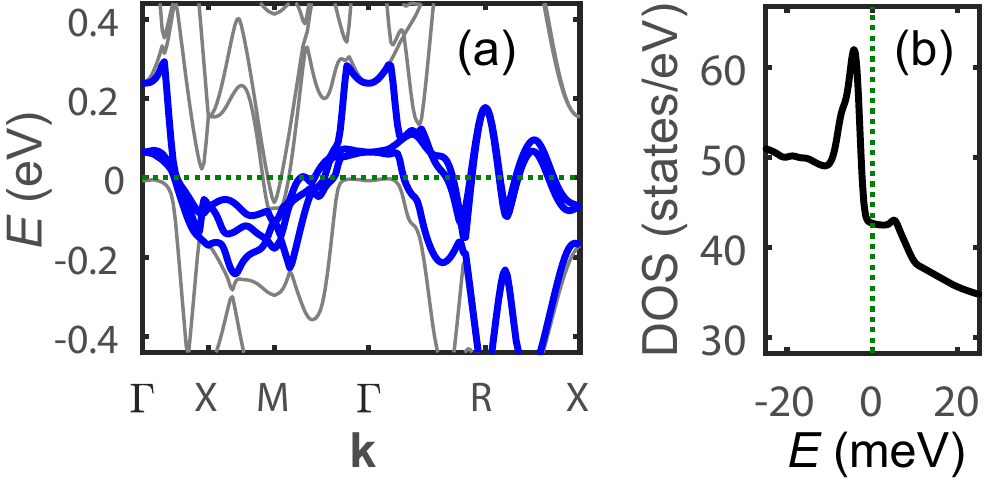}
	\centering
	\caption{\label{Fig:bands} (a) DFT band-structure of Lu$_3$Ir$_4$Ge$_{13}$, showing the bands in the vicinity of the Fermi level along the high-symmetry directions. The blue color indicates the three bands that cross the Fermi level and contribute most to the carrier density. (b) Corresponding density of states (DOS), exhibiting a narrow van Hove singularity about 4 meV below the Fermi level.}
\end{figure}

\begin{figure}[htbp]
	\includegraphics[clip,width=\columnwidth]{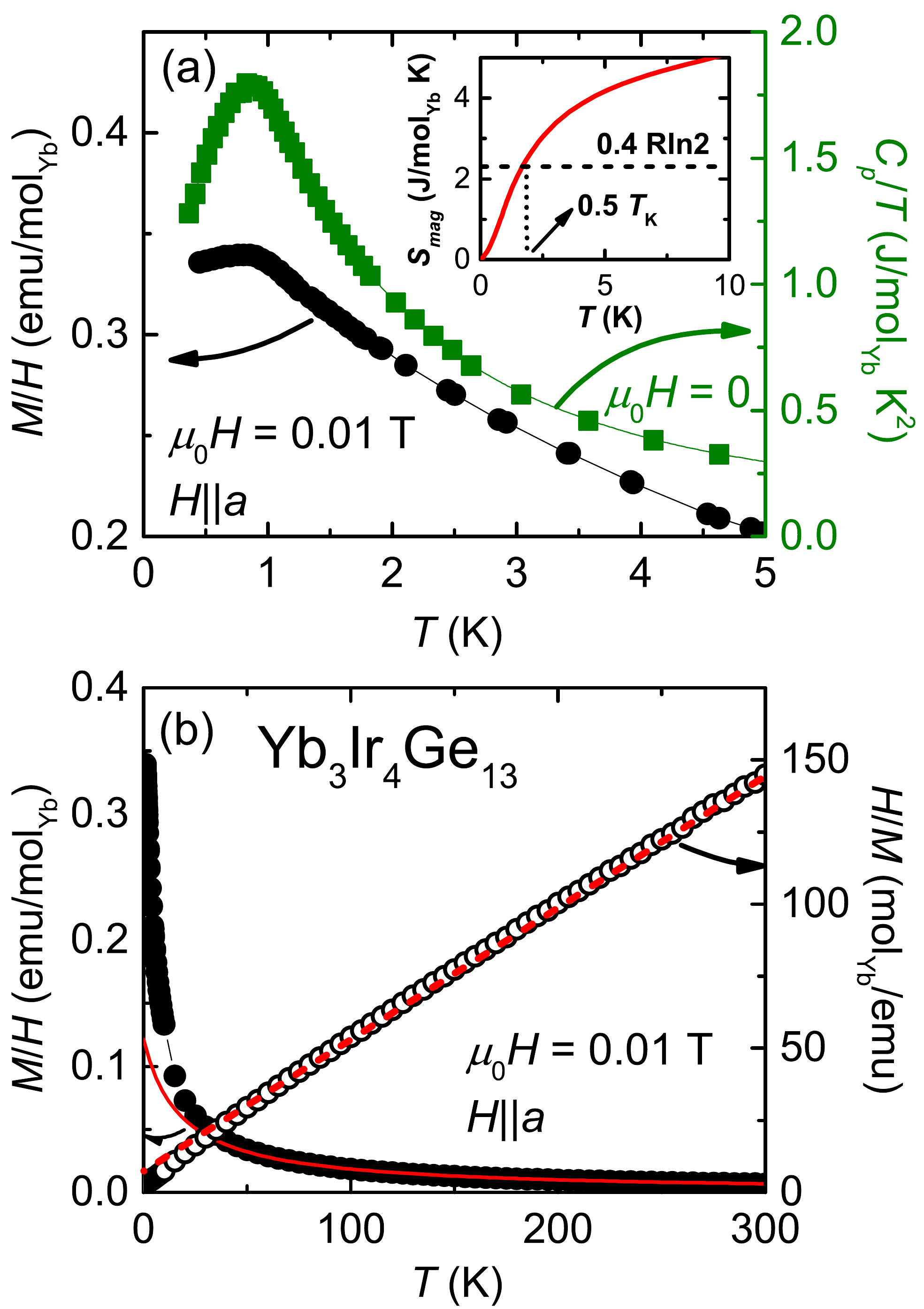}
\caption{\label{Fig3} (a) Low-$T$ $M/H$ vs. $T$ (left, circle) and zero-field specific heat $C_p/T$ vs. $T$ (right, square) of Yb${_3}$Ir${_4}$Ge$_{13}$. The inset shows the temperature dependence of magnetic entropy $S_{mag}$ which is calculated from the 4$f$-moment contribution to the specific heat. (b) Magnetic susceptibility $M/H$ vs. $T$ (left) , where $M$ is the magnetization and $H$ is the magnetic field, and inverse magnetic susceptibility $H/M$ vs. $T$ (right) of Yb${_3}$Ir${_4}$Ge$_{13}$. The solid and dashed lines show a Curie-Weiss fit. Note that 1 emu = 1 G cm$^{3}$ = 10 $^{-3}$ Am$^2$.
}
\end{figure}

In order to substantiate this conclusion, we have performed the density-functional theory (DFT) calculations, focusing on LIG, where the absence of $f$-electron bands near the Fermi level allows to better understand the underlying mechanism for low-carrier density. Shown in Fig.~\ref{Fig:bands}(a) is the band structure plotted along high-symmetry lines in the Brillouin zone, from which it follows that there are a total of 5 pseudo-spin degenerate bands that cross the Fermi level. Of these 5 bands, the most significant ones are two hole-like bands and one electron-like band, whereas the remaining two electron-like bands form tiny electron pockets near the $M$ point and contribute very little to the carrier density (see Fig.~\ref{Fig:bands}(a)). The individual band carrier densities are summarized in Table~S1 in the Supplementary Materials, with the total carrier density equal $1.91\times 10^{21}$~cm$^{-3}$. This value matches well with the density inferred from the optical conductivity, provided the average effective mass $m^{\ast} \approx 5 m_e$. Calculations of the effective masses from DFT are unavailable, given the multiple bands and their non-parabolic nature near the Fermi level. Nevertheless, the semi-quantitative agreement with the optical data  is encouraging. Within the DFT framework, we find that LIG is a fully compensated semimetal, with the equal total number of electron-like and hole-like carriers. The band structure contains a nearly flat band close to the $\Gamma$ point, which nearly touches the Fermi level. This band contributes to a van Hove singularity in the density of states (DOS), at $\approx 4$ meV below the chemical potential, as shown in Fig.~\ref{Fig:bands}(b). This finding indicates that LIG should be very sensitive to hole doping; the DOS at the Fermi level could be increased by about 50\% if the chemical potential were shifted to coincide with the van Hove peak. Such a substantial enhancement of the DOS signals that this system may have a propensity towards magnetic ordering, based on a Stoner criterion-type argument.
 
We have so far demonstrated the low-carrier-density properties of YbIG and LIG, and turn next to the magnetic properties of YbIG. The single-ion Kondo temperature $T_{\rm K}  \approx$~3.5~K for YbIG, which is estimated from the magnetic entropy, \textit{i.e.},  $S_{mag}$(0.5 $T_{\rm K}$) = 0.4 Rln2 (inset of Fig.~\ref{Fig3}(a)), implies that above $T =$ 50~K YbIG can be regarded as a local-moment system. Indeed an effective moment $\mu^{exp}_{eff}$ = 4.2 $\mu_{\rm B}/$mol$_{\rm Yb}$, close to the Yb$^{3+}$ free-ion value of $\mu^{calc}_{eff}$ = 4.54 $\mu_{\rm B}/$mol$_{\rm Yb}$, is obtained from the Curie-Weiss fit of the inverse magnetic susceptibility $H/M$ (open symbols, right axis in Fig.~\ref{Fig3}(b)). The fit results in a Weiss temperature $\theta_{\rm W} = - 18$ K, suggesting antiferromagnetic correlations. Upon further cooling, these antiferromagnetic correlations result in a phase transition-like feature at $T^{\ast}_{mag}=$ 0.9 K as seen from the peaks in $M/H$ (circles, left axis) and $C_{p}/T$ (squares, right axis) shown in Fig.~\ref{Fig3}(a). However, the magnetism appears to be very fragile, as indicated by the neutron diffraction and muon spin relaxation ($\mu$SR) measurements discussed next. While neutron diffraction measurements (not shown) clearly observe the diffraction pattern of the crystal structure, no magnetic peaks were found down to 80~mK. This might reflect the fact that the ordered moment of YbIG below $T^{\ast}_{mag}$ is smaller than the neutron diffraction resolution limit. For this reason, we turned to $\mu$SR measurements on YbIG from 2~K down to 0.1~K. As the sample is cooled below $T^{\ast}_{mag}$ there are no spontaneous oscillations in the decay asymmetry. However, there is a continuous evolution in the asymmetry that is well described by a stretched exponential, where both the relaxation rate and the stretching parameter increase monotonically down to the lowest measured temperature (Supplementary Figure~S3). The increase in the relaxation rate indicates a continuous slowing of the Yb$^{3+}$ spin fluctuations. Thus, it is clear that the $T^{\ast}_{mag}$ feature has a magnetic origin. 

At a first glance, the sister compound Ce$_3$Co$_4$Sn$_{13}$ might appear synonymous with YbIG: specific heat measurements reveal a broad peak at 0.8 K in Ce$_3$Co$_4$Sn$_{13}$ \cite{Cornelius2006,Slebarski2012} below which elastic neutron scattering data do not show signs of long-range magnetic order \cite{Christianson2007}, even though inelastic neutron scattering data reveal antiferromagnetic correlations below 15 K \cite{Iwasa2017}. However, the electrical resistivity data on polycrystalline and single crystalline Ce$_3$Co$_4$Sn$_{13}$ show sample dependence \cite{Thomas2006,Lue2012,Slebarski2012}, and the infrared spectroscopy study on the single crystalline sample indicates good metallic response \cite{Ban2017}, unlike the present evidence for low-carrier semimetallic behavior in YbIG and its non-magnetic analogue LIG. Therefore, the fragile magnetism in YbIG and Ce$_3$Co$_4$Sn$_{13}$ are likely of different origins.

\begin{figure*}[htbp]
	\includegraphics[clip,width=\textwidth]{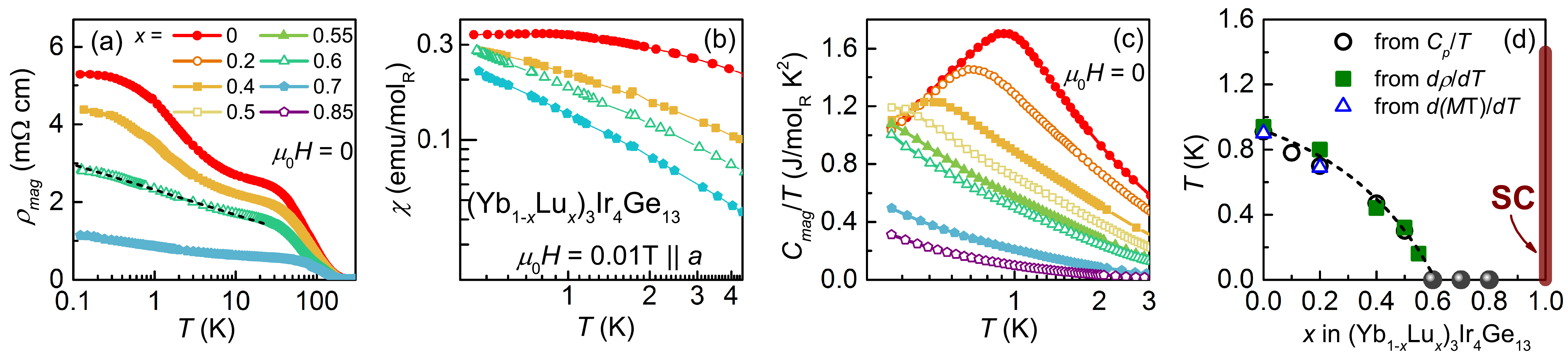}
\caption{\label{Fig4} (a) Zero-field temperature-dependent magnetic electrical resistivity $\rho_{mag}$ of (Yb$_{1-x}$Lu$_x$)${_3}$Ir${_4}$Ge$_{13}$. (b) Low-temperature magnetic susceptibility $M/H$ vs. $T$ at $\mu_0 H$~=~0.01~T$\|a$~axis. (c) Zero-field magnetic specific heat $C_{mag}$/$T$ of (Yb$_{1-x}$Lu$_x$)${_3}$Ir${_4}$Ge$_{13}$ with conduction-electron and phonon contributions being subtracted. (d) $T$ vs. Lu concentration, $x$, phase diagram. The dashed line separates the low-$T$ fragile magnetic state from the high-$T$ paramagnetic state. SC stands for the superconducting state for LIG.}
\end{figure*}

Next we study the substitution series (Yb$_{1-x}$Lu$_x$)${_3}$Ir${_4}$Ge$_{13}$ (YLIG) to further explore the weak magnetism of YbIG. Increasing $x$ gradually reduces $T^{\ast}_{mag}$ and gives rise to nFL-like behavior when the magnetic correlations are suppressed at the critical composition $x_{c}$ = 0.6, as shown in Fig.~\ref{Fig4}. The broad hump around 40 K in resistivity for YbIG due to Kondo effect (Fig.~\ref{Fig1}(a)) is further manifested in the magnetic resistivity $\rho_{mag}$~=~$\rho$(YLIG)$~-~\rho$(LIG) and the hump gradually vanishes with increasing $x$  (Fig.~\ref{Fig4}(a)). At $x_{c}$ = 0.6, $\rho_{mag}$ diverges logarithmically down to the lowest measured temperature, and the divergence persists, even for $x > x_{c}$. Because of the underlying semimetal character, this cannot be directly compared with metallic HFs near quantum critical points (QCPs), where the dependence of $\rho \sim T^{\alpha}$ with $\alpha < 2$ is found \cite{SiSteglich,Loehneysen}. The magnetic susceptibility $M/H$ and the magnetic specific heat, $C_{mag}/T~=~C/T$(YLIG)$~-~C/T$(LIG), both diverge at $x = x_{c}$ upon cooling, as shown in Fig.~\ref{Fig4} (b) and (c), respectively. From $C_{p}/T$, $d\rho/dT$, and $d(MT)/dT$ we summarized $T^{\ast}_{mag}$ values as a function of $x$ in YLIG, as shown in Fig.~\ref{Fig4} (d). The trend of the $T^{\ast}_{mag}$ line is akin to the phase boundary of the generic phase diagram with a quantum phase transition, where the antiferromagnetic transition is continuously tuned to a QCP by a non-thermal control parameter \cite{SiSteglich,Loehneysen}. The divergent behavior of $M/H$ and $C_{mag}/T$ at $x = x_{c}$ in YLIG is similar to that found in other Yb-based systems close to a QCP \cite{Paschen,Steppke}, although the divergence persists for compositions above x$_c$. %\textcolor{red}{could you see Schottky coming in, and therefore no FL because of some power law divergence due to Schottky? Basically, is there a way to rationalize that even above a QCP, you may see something other than FL?}

A recent theoretical model is aimed at describing the dilute-carrier Kondo limit in a honeycomb lattice \cite{Feng and Si}. In this context, the crystallographic details of YbIG become relevant: Given the even number of Yb atoms/sites (forty-eight) in one unit cell (inset of Fig.~\ref{Fig1}(a), and Supplementary Materials), YbIG has the structural framework for which to employ the even-site-per-unit-cell Kondo-lattice model in the dilute-carrier limit. The theoretical study of such a model has shown that the Kondo effect in this regime gives rise to semimetallic behavior \cite{Feng and Si}. Although the theoretical model describes a honeycomb lattice symmetry instead of the tetragonal crystal system observed here, the essential physics of even-site low-carrier framework is unambiguously the same. The $\rho(T)$ behavior we have observed in the low-temperature regime of YbIG is compatible with this mechanism. At the same time, the RKKY interaction in this regime is expected to be long-ranged and, thus, inherently frustrated. This renders the fragile magnetism we have observed in YbIG a rather natural consequence. As such, our work has not only discovered an entirely new regime of Kondo effect and magnetic correlations in YbIG, but also revealed how such effects can be tuned through the Lu-for-Yb substitution in YLIG, leading to singular nFL-like properties.
	
In summary, YbIG is a low-carrier semimetal with Kondo behavior arising from a low-carrier \textit{non-magnetic} reference compound, LIG, and displays fragile magnetism. The complex intertwinement of low-carrier character and
magnetism in YLIG opens up an entirely new venue to explore the strong correlation effects in a semimetallic environment.

We thank B.S. Hitti for assistance with the $\mu$SR measurements, and J. M. Santiago for assistance with the neutron diffraction measurements. We thank K. Isawa for fruitful discussions. B.R., C.-L.H. and E.M. acknowledge support from the Gordon and Betty Moore Foundation EPiQS Initiative through grant GBMF 4417.  N.L.W. is supported by the National Science Foundation of China (No. 11327806) and the National Key Research and Development Program of China (No.2016YFA0300902, 2017YFA0302904). A.M.H., M.N.W., and G.M.L. acknowledge support from NSERC of Canada. The neutron scattering work at Rice is supported by the U.S. DOE, BES under Contract No. DE-SC0012311 (P.D.). A part of the material characterization work at Rice is supported by the Robert A. Welch Foundation Grant No. C-1839 (P.D.). The theoretical work at Rice University was supported by NSF CAREER Grant No.~DMR-1350237 (A.H.N.) and the Robert A.\ Welch Foundation Grant No.\ C-1818 (V.L. and A.H.N.); as well as by the NSF Grant No.\ DMR-1611392 and the Robert A.\ Welch Foundation Grant No.\ C-1411 (Q.S.), with travel support provided by the ARO Grant No.\ W911NF-14-1-0525. UT Dallas acknowledges support from NSF DMR-1700030. G.M.L. acknowledges support from the Natural Sciences and Engineering Research Council (Canada). S.D. acknowledges Junior Short Award – ICAM-I2CAM QuantEmX scientific report. S.D. and S.P. gratefully acknowledge financial support from the Austrian Science Funds (FWF doctoral program W1243 and FWF I2535-N27) and the U.S. Army research office (grant W911NF-14-1-0496).  The identification of any commercial product or trade name does not imply endorsement or recommendation by the National Institute of Standards and Technology.

\end{document}